\documentclass[12pt]{article}

\usepackage{amsmath}
\usepackage{amssymb}
\usepackage{mathrsfs}
\usepackage{graphicx}
\usepackage{enumerate}

\usepackage[margin=1in]{geometry}

\usepackage[pdftex,colorlinks=true,linkcolor=blue,citecolor=blue,urlcolor=blue]{hyperref}

\begin{document}

\title{Divergences in Holographic Complexity}
\author{Alan Reynolds\footnote{a.p.reynolds@durham.ac.uk}\,\,  and Simon F. Ross\footnote{s.f.ross@durham.ac.uk} \\  \bigskip \\ Centre for Particle Theory, Department of Mathematical Sciences \\ Durham University\\ South Road, Durham DH1 3LE}

\maketitle

\abstract{We study the UV divergences in the action of the ``Wheeler-de Witt patch'' in asymptotically AdS spacetimes, which has been conjectured to be dual to the computational complexity of the state of the dual field theory on a spatial slice of the boundary. We show that including a surface term in the action on the null boundaries which ensures invariance under coordinate transformations has the additional virtue of removing a stronger than expected divergence, making the leading divergence proportional to the proper volume of the boundary spatial slice. We compare the divergences in the action to divergences in the volume of a maximal spatial slice in the bulk, finding that the qualitative structure is the same, but subleading divergences have different relative coefficients in the two cases.}
 
\clearpage 

\section{Introduction}

In principle, holography provides a well-defined non-perturbative formulation of quantum gravity. But to really use it to address questions about the nature of spacetime, we need to understand the emergence of the bulk spacetime from the dual field theory description. In \cite{Susskind:2014rva}, Susskind conjectured a new relation between the bulk geometry and the complexity of the dual boundary state. The quantum computational complexity is a measure of the minumum number of elementary gates needed in a quantum circuit which constructs a given state  starting from a specified simple reference state (see e.g. \cite{Osborne:c}). 
This proposal was refined  in \cite{Stanford:2014jda} into the conjecture that the computational complexity of the boundary state at a given time (on some spacelike slice of the boundary) could be identified with the volume of a maximal volume spacelike slice in the bulk, ending on the given boundary slice. This will be referred to as the CV conjecture. This was further developed in \cite{Susskind:2014jwa,Susskind:2014moa}. 

More recently, it was conjectured that the complexity is related instead to the action of a Wheeler-de Witt patch in the bulk bounded by the given spacelike surface \cite{Brown:2015bva,Brown:2015lvg}. This is referred to as the CA conjecture. An appropriate prescription for calculating the action for a region of spacetime bounded by null surfaces was obtained in \cite{Lehner:2016vdi}. Further related work is \cite{Barbon:2015ria,Brown:2016wib,Chapman:2016hwi}.
 
 The evidence supporting this conjecture comes so far from the study of black hole spacetimes. Both the CV and CA conjectures produce results for the complexity that grow linearly in time, with 
 \begin{equation} \label{ctd}
 \frac{d \mathcal C}{dt} \propto M. 
 \end{equation}
This is consistent with general expectations for the behaviour of the complexity for excited states in the field theory. In these investigations, questions about the UV structure of the complexity were avoided, as the contributions from the asymptotic region of the spacetime cancel out in considering the time derivative.\footnote{Another way to cancel UV contributions is to consider the difference between two spacetimes with the same asymptotic structure, as in \cite{Chapman:2016hwi}.}

However, it is interesting to understand the divergences in the holographic complexity. In both the CV and CA conjectures there will be UV divergences, as the volume or action of the spacetime region in the bulk is divergent near the boundary. We would expect that as for the holographic entanglement entropy \cite{Ryu:2006bv,Ryu:2006ef}, these divergences are physical, signalling divergent contributions to the complexity associated with the UV degrees of freedom in the field theory. For the entanglement entropy, the leading divergence is proportional to the area of the entangling surface, and this can be understood as reflecting entanglement of UV modes across this boundary \cite{Bombelli:1986rw,Srednicki:1993im}. While a detailed understanding of the divergences of the complexity from the field theory perspective does not yet exist, we can study the divergences in the holographic calculation, and see if they have a reasonable form. It is also interesting to compare the divergences between the CV and CA prescriptions, and see to what extent they compute different versions of complexity. 

While this paper was in preparation, a preprint appeared studying these divergences \cite{Carmi:2016wjl}. The purpose of the present note is to add a simple observation to that work. There is a term identified in \cite{Lehner:2016vdi} which can be added to the action which cancels a coordinate-dependence in that prescription. If we add this contribution, it cancels the leading divergence in the CA prescription, so that the divergence structure of this action is the same as in the CV prescription. The leading divergence in both cases is proportional to the volume of the boundary time slice. Such a divergence appears reasonable from a field theory perspective. Considering subleading contributions, we find that in both cases they can be expressed in terms of the geometry of the slice, but the CV and CA prescriptions differ. 

In section \ref{review}, we review the CV and CA conjectures, and their application to the black hole examples. We discuss the coordinate-dependence of the action proposed in \cite{Lehner:2016vdi}, and introduce the term cancelling it. In section \ref{div}, we consider the UV divergences in the CV and CA calculations, and show that including this term cancels the leading divergence in the CA calculation. We consider subleading contributions and show that they have similar structures, depending on local geometric invariants of the boundary geometry, but note that the two prescriptions will differ in general. We study the computation on global AdS to illustrate this difference. 

\section{Review of CV and CA}
\label{review}

In the CV conjecture of \cite{Susskind:2014rva}, the complexity $\mathcal C$ of the state $|\Psi \rangle$ of a holographic field theory on some spatial slice $\Sigma$ on the boundary of an asymptotically AdS spacetime is identified with the volume $V$ of the maximal volume codimension one slice $B$ in the bulk having its boundary on $\Sigma$,
\begin{equation}
{\mathcal C_V} = \frac{V(B)}{G_N l_{AdS}},
\end{equation}
This has a UV divergence proportional to the volume of $\Sigma$. If we interpret this as part of the physical complexity, it could be interpreted as reflecting the operations required to set up the appropriate short-distance structure of the state $|\Psi \rangle$  starting from some reference state. Qualitatively, this is reasonable; if we imagine modelling the field theory as a lattice, the reference state could be a simple product state on the lattice sites. A Hadamard state in the field theory will not have such a product structure; the absence of high energy excitations implies short-range entanglement/correlation in the state. Setting up this entangled state from the reference product state would require a number of elementary operations which will grow proportional to the volume of the field theory.  
 
 In \cite{Brown:2015bva,Brown:2015lvg}, an alternative CA conjecture was proposed. This identifies the complexity of $|\Psi \rangle$ with the action of the ``Wheeler-de Witt patch", the domain of development of the slice $B$ considered previously. The proposal is that
\begin{equation}
{\mathcal C_A} = \frac{S_W}{\pi \hbar},
\end{equation}
where $S_W$ is the action of the Wheeler-de Witt patch. This proposal has the advantage that the formula is more universal, containing no explicit reference to a bulk length scale. It also turns out to be easier to calculate, as we don't have a maximisation problem to solve. Finding the Wheeler-de Witt patch for a given boundary slice is easier than finding the maximal volume slice. 

The Wheeler-de Witt patch has null boundaries, for which the appropriate boundary terms needed for the Einstein-Hilbert action were not yet known. In \cite{Lehner:2016vdi}, inspired by the CA conjecture, a prescription for the action of a region of spacetime containing null boundaries was constructed (see also \cite{Parattu:2015gga,Jubb:2016qzt}). The prescription was obtained by requiring that the variation of the action vanish on-shell when the variation of the metric vanishes on the boundary of the region. The resulting form for the action is 
\begin{eqnarray} \label{watact}
S_{\mathcal V} &=& \int_{\mathcal V} (R- 2 \Lambda) \sqrt{-g} dV + 2 \sum_{T_i} \int_{T_i} K d\Sigma + 2 \sum_{S_i} \mathrm{sign}(S_i) \int_{S_i} K d \Sigma \\ \nonumber
&& - 2 \sum_{N_i} \mathrm{sign}(N_i) \int_{N_i} \kappa dS d\lambda + 2 \sum_{j_i} \mathrm{sign}(j_i) \oint \eta_{j_i} dS + 2 \sum_{m_i} \mathrm{sign}(m_i) \oint a_{m_i} dS.
\end{eqnarray}
In this expression
\begin{itemize}
\item $T_i$ and $S_i$ are respectively timelike and spacelike components of the boundary of the region $\mathcal V$, and $K$ is the trace of the extrinsic curvature of the boundary. For $T_i$, the normal is taken outward-directed from $\mathcal V$. For $S_i$, the normal is always taken future-directed, and sign$(S_i) = 1(-1)$ if $\mathcal V$ lies to the future (past) of $S_i$, that is if the normal vector points into (out of) the region of interest. 
\item $N_i$ are null components of the boundary of $\mathcal V$, $\lambda$ is a parameter on null generators of $N_i$, increasing to the future, $dS$ is an area element on the cross-sections of constant $\lambda$, and $k^\alpha \nabla_\alpha k^\beta = \kappa k^\beta$, where $k^\alpha= \partial x^\alpha/\partial \lambda$ is the tangent to the generators. sign$(N_i) = 1(-1)$ if $N_i$ lies to the future (past) of   ${\mathcal V}$. 
\item $j_i$ are junctions between non-null boundary components,  where $\eta$ is the logarithm of the dot product of normals. We do not give the rules in detail as such junctions do not occur for Wheeler-de Witt patches; see \cite{Lehner:2016vdi} for full detail.
\item $m_i$ are junctions where one or both of the boundary components are null. We have a null surface with future-directed tangent $k^\alpha$ and either a spacelike surface with future directed unit normal $n^\alpha$, a timelike surface with outward directed unit normal $s^\alpha$, or another null surface with future-directed tangent $\bar k^\alpha$, and
\begin{equation}
a = \left\{ \begin{array}{c} \ln | k \cdot n| \\ \ln| k \cdot s| \\ \ln|k \cdot \bar k/2| \end{array} \right.
\end{equation}
respectively. sign$(m_i) = +1$ if $\mathcal V$ lies to the future (past) of the null boundary component and $m_i$ is at the past (future) end of the null component, and sign$(m_i) = -1$ otherwise.
\end{itemize}

While this action is diffeomorphism invariant under changes of coordinates in the bulk and on the timelike and spacelike boundaries, \cite{Lehner:2016vdi} show that it depends on the choice of coordinate $\lambda$ on the null boundary components. This coordinate dependence seems a highly undesirable feature. Coordinate independence on the timelike and spacelike boundaries was incorporated as an assumption in constructing the form of the action. This was built in, as it was possible to work with covariant tensors throughout the calculation. On the null boundaries, such a manifestly coordinate independent formalism does not seem to exist, but one would still like to require that the final expression exhibit coordinate independence as a fundamental feature. Fortunately,  \cite{Lehner:2016vdi}  found that the coordinate dependence could be eliminated by adding to the action a term
\begin{equation} \label{acorr}
\Delta S = -2 \sum_{N_i}  \mathrm{sign}(N_i) \int_{N_i} \Theta \ln |\ell \Theta| dS d\lambda, 
\end{equation}
where $\Theta$ is the expansion of the null generators of $N_i$, 
\begin{equation}
\Theta = \frac{1}{\sqrt{\gamma}} \frac{\partial \sqrt{\gamma}}{\partial \lambda},
\end{equation}
where $\gamma$ is the metric on the cross-sections of constant $\lambda$. We will henceforth adopt the action $S = S_{\mathcal V} + \Delta S$ as our definition of the action for a region with null boundaries. 

There is a further ambiguity noted in \cite{Lehner:2016vdi}, which is the freedom to add an arbitrary function independent of the bulk metric to $a_{m_i}$. We see this as a subcase of a general freedom in the action: the requirement that the variation of the action vanish fixes the form of the boundary terms only up to contributions whose variations vanish under variations of the metric. Since the metric variation vanishes on the boundary, this includes the freedom to add arbitrary functions of the intrinsic geometry of the boundary.  

If we ignored the requirement of coordinate independence, this freedom would include the freedom to add terms like \eqref{acorr}, as its variation under metric variations (with the metric fixed on the boundary) vanishes. Since we want to insist on coordinate independence, the coefficient in \eqref{acorr} is fixed, but we still have freedom to add terms which are scalars on the null boundary, such as $ \int_{N_i} \Theta f(\gamma) dS d\lambda$, where $f(\gamma)$ is any scalar function of the cross-section metric $\gamma$ and curvature invariants built from this metric such as its Ricci scalar. Also we have the freedom to add such scalar terms at the corners. 

\section{UV divergences}
\label{div} 

\begin{figure}
\centering 
\includegraphics[width=4cm]{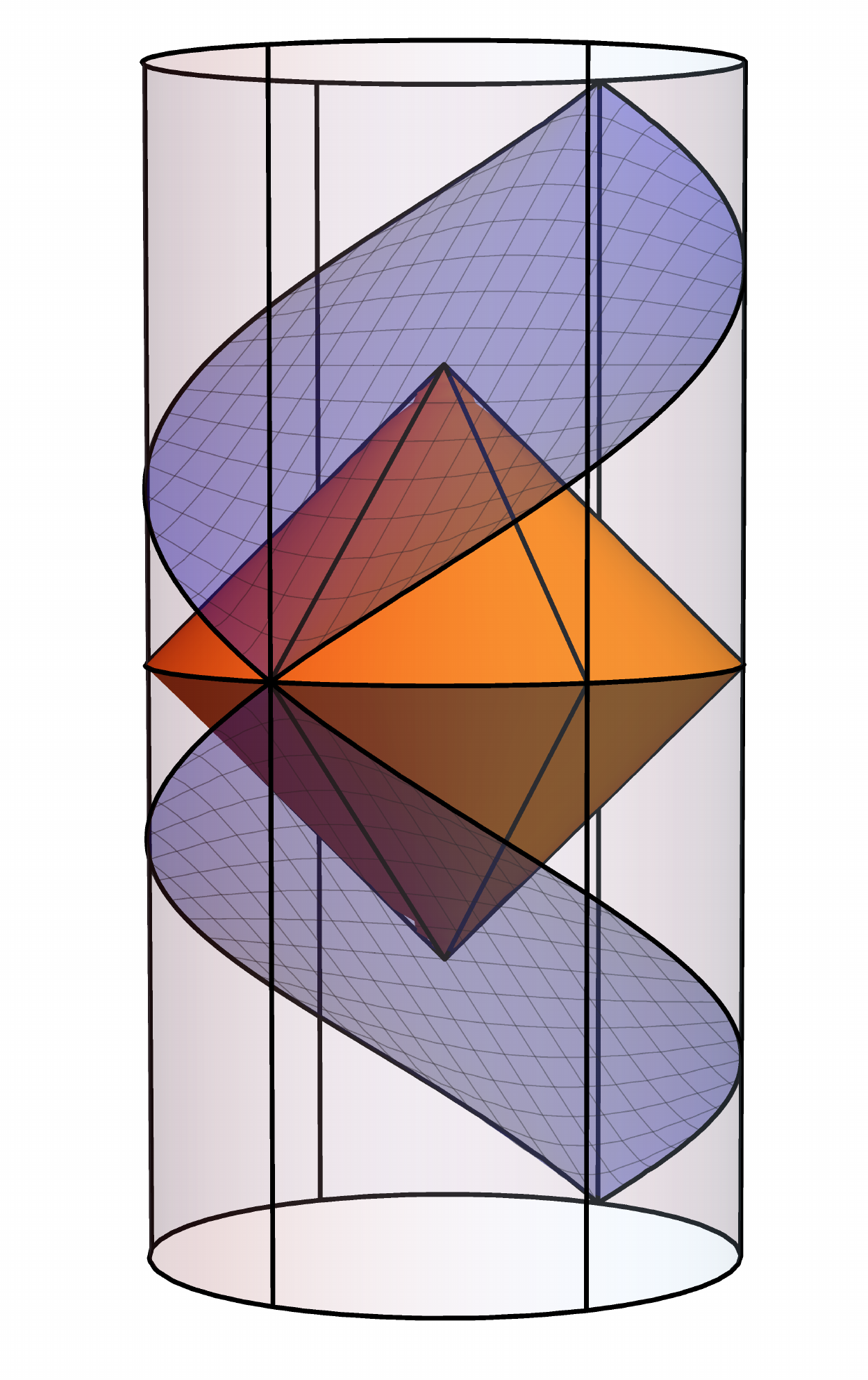}
\caption{AdS, showing the region covered by Poincare coordinates and the Wheeler-de Witt patch of the $t=0$ surface.} \label{pads}
\end{figure}

We now turn to the consideration of the UV divergences in the action for the Wheeler-de Witt patch. The simplest case to consider is AdS$_{d+1}$ in Poincare coordinates,
\begin{equation} \label{adsp}
ds^2 = \frac{\ell^2}{z^2} ( dz^2 - dt^2 + d\vec{x}^2),
\end{equation}
which is dual to the field theory in flat space. We consider a $d+1$ dimensional AdS space, with a $d$ dimensional boundary. If we ask for the complexity of the field theory on the $t=0$ surface, cut off at $z = \epsilon$, the Wheeler de Witt patch lies between $t = z-\epsilon$ and $t = -(z-\epsilon)$. Note that although these coordinates do not cover the full spacetime, the Wheeler-de Witt patch lies inside the region covered by this coordinate patch, as shown in figure \ref{pads}, so we can calculate its action in these coordinates.

For the CV conjecture, the maximal volume slice with boundary at $t=0$ is simply the $t=0$ surface in the bulk, whose volume is 
\begin{equation} \label{vol}
V(B) = \int dz d^{d-1} x \sqrt{h} = \ell^d V_x  \int_\epsilon^\infty \frac{dz}{z^d} = \frac{\ell^d V_x}{(d-1) \epsilon^{d-1}},
\end{equation}
where $V_x$ is the IR divergent coordinate volume in the $\vec x$ directions. Thus, the complexity calculated according to the CV prescription is, up to an overall constant, 
\begin{equation}  \label{cvol}
\mathcal{C}_V =  \frac{\ell^{d-1} V_x}{(d-1) G_N \epsilon^{d-1}}.  
\end{equation}
This is proportional to the volume of the space the field theory lives in, in units of the cutoff. This has both an IR and a UV divergence, which is physically reasonable if we think of the complexity as defined with respect to some product lattice state, as previously discussed. 

Turning to the CA conjecture, consider the Wheeler-de Witt patch of this cutoff surface.\footnote{We could alternatively take the original Wheeler-de Witt patch of the surface at $t=0,z=0$ and cut off the corner at $z= \epsilon$, producing a small timelike boundary component. This would produce a different set of coefficients for subleading divergences \cite{Carmi:2016wjl}.} The action of the Wheeler-de Witt patch with the prescription of \cite{Lehner:2016vdi} is 
\begin{equation} \label{adsa}
S_W = \int_{W} (R- 2 \Lambda) \sqrt{-g} dV  - 2  \int_{F} \kappa dS d\lambda   + 2  \int_{P} \kappa dS d\lambda  -  2   \oint_{\Sigma} a dS, 
\end{equation}
where $F$ ($P$) is the future (past) null boundary of the Wheeler-de Witt patch, and $\Sigma$ is the surface at $t=0$, $z=\epsilon$. The light cones of the boundary surface are at $t = \pm (z- \epsilon)$, and $R - 2 \Lambda = -2d/\ell^2$, so the volume integral is 
\begin{equation}
S_{\mathit{Vol}} = - 2 \frac{d}{\ell^2} \int_\epsilon^\infty dz \int_{-(z-\epsilon)}^{z- \epsilon} dt \frac{\ell^{d+1}}{z^{d+1}} V_x = -4 \frac{\ell^{d-1} V_x}{(d-1) \epsilon^{d-1}}.
\end{equation}
This has a very similar structure to the volume in \eqref{vol}, but this term is negative, so it is clearly important to include the boundary contributions identified in \cite{Lehner:2016vdi} to obtain a sensible result for the complexity. 

In calculating \eqref{adsa}, it is convenient to adopt an affine parametrization of the null surfaces, so that the integrals over the future and past boundaries do not contribute. Let us take the affine parameters along the null surfaces to be 
\begin{equation} \label{afp}
\lambda = - \frac{\ell^2}{\alpha z} \mbox{ on F}, \quad \lambda = \frac{\ell^2}{\beta z} \mbox{ on P}, 
\end{equation}
where we introduce the arbitrary constants $\alpha, \beta$ to exhibit explicitly the remaining coordinate dependence. This gives $k = \alpha z^2/\ell^2 (\partial_t + \partial_z)$, $\bar k = \beta z^2/\ell^2 (\partial_t - \partial_z)$.  The boundary corner term is thus
\begin{equation}
S_{\Sigma} = - 2 \frac{\ell^{d-1} V_x}{\epsilon^{d-1}} \ln(\alpha \beta \epsilon^2/\ell^2).  
\end{equation}
Thus, the action calculated according to \eqref{adsa} is 
\begin{equation}
S_W = \frac{\ell^{d-1} V_x}{\epsilon^{d-2}}[ - 4 \ln( \epsilon/\ell) - 2 \ln(\alpha\beta) - \frac{1}{d-1} ].  
\end{equation}
This has two undesirable features: it depends on the normalization $\alpha$, $\beta$ of the affine parameters on the two null surfaces, and it diverges like $\epsilon^{-(d-2)} \ln \epsilon$, which is faster than the volume of the space the field theory lives in. These effects drop out if we consider the time-dependence as in \eqref{ctd}, but they are both problematic if we want to consider the action as dual to the actual complexity of the state. The first implies that the identification will require some choice of normalization for the affine parameters, which seems strange; these are just coordinates and should have no physical content. The second implies the complexity would have a stronger than volume divergence, which seems not so easy to understand in terms of  a simple lattice model. 

Fortunately, both these problems are removed once we include the additional contribution \eqref{acorr}. The metric on F has $\sqrt{\gamma} = \ell^{d-1}/z^{d-1}$, so the expansion is
\begin{equation}
\Theta = \frac{1}{\sqrt{\gamma}} \frac{\partial \sqrt{\gamma}}{\partial \lambda} = - \frac{1}{\sqrt{\gamma}} \alpha  \frac{z^2}{\ell^2}  \frac{\partial \sqrt{\gamma}}{\partial z} = (d-1) \alpha \frac{z}{\ell^2}, 
\end{equation}
so the surface term is
\begin{eqnarray}
S_{F} &=& -2 (d-1) \ell^{d-1} V_x \int z^{-(d-2)} \ln (\alpha (d-1) z/\ell) \alpha d\lambda \\ &=&  2 (d-1) \ell^{d-1} V_x \int_\epsilon^\infty z^{-d} \ln (\alpha (d-1) z/\ell) dz \nonumber \\  &=&  2\frac{ \ell^{d-1}}{\epsilon^{d-1}} V_x \left( \ln( \alpha (d-1) \epsilon/\ell) + \frac{1}{d-1} \right),  \nonumber 
\end{eqnarray}
and $S_P = 2\frac{ \ell^{d-1}}{\epsilon^{d-1}} V_x \left( \ln(\beta (d-1) \epsilon/\ell) + \frac{1}{d-1} \right)$, so 
\begin{equation}
S = S_W + \Delta S = S_{\mathit{Vol}} + S_\Sigma + S_F + S_P = 4  \frac{ \ell^{d-1}}{\epsilon^{d-1}} V_x \ln(d-1).  
\end{equation}
The dependence on $\alpha$, $\beta$ cancels out by construction, as the additional terms were introduced to eliminate the coordinate dependence in \eqref{watact}. The surprise is that this also leads to the cancellation of the logarithmic divergence.\footnote{For the case $d=1$, that is AdS$_2$, the null surfaces are one-dimensional, and there is no expansion, so we cannot define a term analogous to \eqref{acorr} to cancel the logarithmic divergence. In this case the CV calculation is also logarithmically divergent. It would be interesting to understand this better, as this case will emerge if we want to apply these complexity ideas to near-horizon geometries of near-extremal black holes.}  This provides a strong additional support for the idea that \eqref{acorr} should be included in the calculation of the action. The result now has the same structure as that obtained in the CV calculation \eqref{cvol}; since we do not understand the relation between the complexity and spacetime very precisely, the difference in the overall coefficient is not particularly significant.

\subsection{Subleading contributions}

If we consider asymptotically AdS spacetimes, there will also be subleading divergences. It is interesting to consider these contributions and investigate whether the cancellation of the leading logarithmic divergence in \eqref{watact} obtained on adding \eqref{acorr} extends to subleading terms. It is also interesting to compare the structure of divergences in the CV and CA calculations. 

We consider an asymptotically AdS$_{d+1}$ solution of the vacuum Einstein equations. The metric in the asymptotic region can then be written in the Fefferman-Graham gauge \cite{Fefferman:2007rka,Graham:1999pm}
\begin{equation} \label{gmet} 
ds^2 = \frac{\ell^2}{z^2} (dz^2 + g_{\mu\nu}(x^\mu,z) dx^\mu dx^\nu), 
\end{equation}
where the metric along the boundary directions has a power series expansion in $z$, 
\begin{equation} 
g_{\mu\nu}(x^\mu,z) = g_{\mu\nu}^{(0)}(x^\mu) + z^2 g_{\mu\nu}^{(1)}(x^\mu) + \ldots. 
\end{equation}

We can give a simple general argument which shows that the cancellation of the leading logarithmic term extends to all the terms of the form $\epsilon^{-n} \log \epsilon$. Logarithmic divergences come from the corner contribution, 
\begin{equation} 
S_\Sigma = -  2  \oint_{\Sigma} \ln| k \cdot \bar k/2| \sqrt{\gamma} d^{d-1} \sigma,  
\end{equation}
and from the additional contribution on the two null surfaces. Considering the future,
\begin{equation} 
S_F = -2 \int \Theta \ln |\ell \Theta| \sqrt{\gamma} d^{d-1}x d\lambda, 
\end{equation}
Now using the fact that the expansion is $\Theta = \frac{1}{\sqrt{\gamma}} \partial \sqrt{\gamma}/\partial \lambda$, we can rewrite this as 
\begin{equation}
S_F = -2  \int \partial_\lambda \sqrt{\gamma}  \ln |\ell \Theta| d^{d-1}x d\lambda, 
\end{equation}
and integrate by parts on $\lambda$. Since $\Sigma$ is a past endpoint of the future surface, we obtain 
\begin{equation} \label{acorr2}
S_F = 2   \oint_{\Sigma} \sqrt{\gamma}  \ln |\ell \Theta| d^{d-1}\sigma  +2 \int \sqrt{\gamma}  \frac{ \partial_\lambda \Theta}{\Theta} d^{d-1}x d\lambda, 
\end{equation}
dropping a boundary term at the other boundary of the null surface which is irrelevant to the asymptotic calculation. The second term will only contribute power-law divergences, so the logarithmic divergences will come solely from the integral over $\Sigma$. Note also that it is this integral over $\Sigma$ which cancels the coordinate-dependence in \eqref{watact}; the second term is coordinate-independent.\footnote{In fact, one could take an alternative prescription for resolving the issues in \eqref{watact} where one just added the first term in \eqref{acorr2}, rather than the whole expression \eqref{acorr}.} There is a similar contribution from the past surface; 
\begin{equation}
S_P =  2  \int \partial_\lambda \sqrt{\gamma}  \ln |\ell \Theta| d^{d-1}x d\lambda, 
\end{equation}
and the boundary term has the opposite sign because $\Sigma$ is a future boundary of the past surface, so
\begin{equation} 
S_P = 2   \oint_{\Sigma} \sqrt{\gamma}  \ln |\ell \Theta| d^{d-1}\sigma  -2 \int \sqrt{\gamma}  \frac{ \partial_\lambda \Theta}{\Theta} d^{d-1}x d\lambda.
\end{equation}
The logarithmic divergences in the full action are then contained in the terms involving integrals on $\Sigma$, 
\begin{equation} 
S = \ldots  + 2  \oint_{\Sigma} (\ln |\ell \Theta_F| +  \ln |\ell \Theta_P| - \ln| k \cdot \bar k/2|) \sqrt{\gamma} d^{d-1} \sigma  
\end{equation}
 We know from the previous calculation that the leading order logarithmic term cancels. Subleading terms coming from the expansion of $\sqrt{\gamma}$ will then also cancel. Subleading terms in the argument of the logarithm will give power law divergences, once we expand $\ln (\epsilon + B\epsilon^2 + \ldots) = \ln \epsilon + \ln (1 + B \epsilon + \ldots) \approx \ln \epsilon + B\epsilon + \ldots$. Thus, there are no subleading terms of the form $\epsilon^{-n} \ln \epsilon$; once we include \eqref{acorr} the divergences are a power series expansion in $\epsilon$.\footnote{This argument is valid for all the terms of the form $\epsilon^{-n} \log \epsilon$ for $n>0$; once we reach the order in the Fefferman-Graham expansion where we encounter the free data in the asymptotic expansion, there may be contributions to either CV or CA calculations at order $\log \epsilon$.} 

We will now extend the explicit calculation of the first subleading corrections to the action in \cite{Carmi:2016wjl} to include the additional contribution along the null surfaces. We will see explicitly that the logarithmic terms cancel, as predicted by the general argument above. We assume we are in $d > 2$ where the term of order $z^2$ is determined locally by the boundary metric $g_{\mu\nu}^{(0)}$, \cite{deHaro:2000vlm,Skenderis:2002wp}
\begin{equation} 
g_{\mu\nu}^{(1)} (x^\mu) = - \frac{\ell^2}{(d-2)} \left( R_{\mu\nu}[ g^{(0)}] - \frac{g^{(0)}_{\mu\nu}}{2(d-1)} R[g^{(0)}] \right),  
\end{equation}
where $R_{\mu\nu}$ and $R$ are the Ricci tensor and Ricci scalar for the boundary metric.  It should be straightforward to extend the analysis to further subleading orders, but we will see interesting differences already at the first subleading order. 

We consider a boundary slice at $t=0$, in the cutoff surface at $z= \epsilon$, and calculate subleading divergences in the complexity. As in \cite{Carmi:2016wjl}, we restrict consideration to cases where the boundary metric is 
\begin{equation} \label{bmet}
g_{\mu\nu}^{(0)} dx^\mu dx^nu = - dt^2 + h_{ab} (t,\sigma^a) d\sigma^a d \sigma^b. 
\end{equation}
This is general enough to include many cases of interest, and considerably simplifies the determination of the Wheeler-de Witt patch. 

In \cite{Carmi:2016wjl}, the subleading contributions to the volume of the maximal volume slice were determined, finding
\begin{equation} \label{slcv} 
\mathcal{C}_V = \frac{\ell^{d-1}}{(d-1) G_N \epsilon^{d-1}} \int d^{d-1} \sigma \sqrt{h} \left[ 1 - \frac{d-1}{2(d-2)(d-3)} \epsilon^2 \left( R_a^a - \frac{1}{2} R - \frac{(d-2)^2}{(d-1)^2} K^2 \right) + \ldots \right], 
\end{equation}
where $h$ is the determinant of the metric $h_{ab}$ in \eqref{bmet} at $t=0$,  $R^a_a = h^{ab} R_{ab}$ is the trace of the projection of the boundary Ricci tensor into the $t=0$ surface, and $K$ is the trace of the extrinsic curvature of the $t=0$ surface in the boundary metric \eqref{bmet}. Thus, the first subleading divergence can be expressed in terms of local geometric features of the boundary metric.  The first subleading contributions to the action \eqref{watact} were also evaluated in \cite{Carmi:2016wjl}, obtaining
\begin{eqnarray} \label{carmia}
\mathcal C_A(S_W) &=& - \frac{\ell^{d-1}}{4\pi^2 G_N (d-1) \epsilon^{d-1}} \int d^{d-1} \sigma \sqrt{h} \left[ 1 \phantom{\frac{1}{2}}  \right. \\ && \left. + \frac{\epsilon^2}{4 (d-2)(d-3)} \left( 4 K^2 + 4 K_{ab} K^{ab} + (d-7) R - 2 (d-3) R_a^a \right)  \right] \nonumber \\ && + \frac{\ell^{d-1}}{4\pi^2 G_N \epsilon^{d-1}} \log\left( \frac{\ell}{\sqrt{\alpha \beta} \epsilon} \right) \int d^{d-1} \sigma \sqrt{h} \left[ 1 - \epsilon^2 \frac{1}{2(d-2)} (R^a_a - \frac{1}{2} R )  \right] + \ldots \nonumber
\end{eqnarray}
We want to consider the effect of adding \eqref{acorr}.

A key feature of the calculation in \cite{Carmi:2016wjl} is that the assumption that the boundary metric has the form \eqref{bmet} implies that at first subleading order, the tangents to the null generators take the form 
\begin{equation} 
k = \frac{\alpha}{\ell^2} (z^2 \partial_z + k^t \partial_t), \quad \bar k = \frac{\beta}{\ell^2} (-z^2 \partial_z + k^t \partial_t),   
\end{equation}
where $k^t$ is determined by requiring these to be null vectors, $k^\mu k_\mu =0$, which gives $k^t = z^2 (g_{tt})^{-1/2}$. This implies that the form of the affine parameter in \eqref{afp} is unchanged to first subleading order. 

Near the boundary, the induced metric on the surfaces of constant $\lambda$ in the null surfaces is thus  $\gamma_{ab} = z^{-2} h_{ab} + g^{(1)}_{ab} + \ldots$. Following \cite{Carmi:2016wjl}, we write 
\begin{equation} 
\sqrt{\gamma} = \frac{\ell^{d-1}}{z^{d-1}} \sqrt{h} ( [1 + q_0^{(2)} z^2 + \ldots ] + [q_1^{(0)} + \ldots] t + [q_2^{(0)} + \ldots] t^2 + \ldots),   
\end{equation}
keeping the first terms in an expansion for small $z$ and $t$, where $h$ is the determinant of $h_{ab}(\sigma^a, t=0)$. Along the null surface $t = (z-\epsilon) + O(z^3)$, so 
\begin{equation} 
\sqrt{\gamma} = \frac{\ell^{d-1}}{z^{d-1}} \sqrt{h} ( 1 + q_1^{(0)} (z-\epsilon) + q_0^{(2)} z^2 + q_2^{(0)}  (z-\epsilon)^2 + \ldots ),   
\end{equation}
\begin{eqnarray} 
\partial_z \sqrt{\gamma} &=& -\frac{\ell^{d-1}}{z^{d}} \sqrt{h} ( (d-1) + q_1^{(0)} ((d-2) z- (d-1)\epsilon) + q_0^{(2)} (d-3) z^2 \\ &&+ q_2^{(0)}((d-3)z^2 -2(d-2) z \epsilon + (d-1) \epsilon^2) + \ldots ),   \nonumber
\end{eqnarray}
so the expansion is 
\begin{equation} 
\Theta = \frac{\alpha z}{\ell^2} \left[ (d-1)  - q_1^{(0)} z - 2 q_0^{(2)} z^2 -2 q_2^{(0)} z(z-\epsilon) + q_1^{(0)2} z (z-\epsilon) + \ldots \right] 
\end{equation}
Performing the integral over $z$, one finds
\begin{eqnarray} 
S_F &=& 2 \frac{\ell^{d-1}}{\epsilon^{d-1}} \int_\Sigma d^{d-1} \sigma \sqrt{h} \left[ \ln (\alpha (d-1)  \epsilon/\ell) ( 1 + q_0^{(2)} \epsilon^2) \right. \\ && \left. +\frac{1}{(d-1)} \left( 1 - q_1^{(0)} \epsilon - q_0^{(2)} \frac{(d-1)}{(d-3)} \epsilon^2 - 2 \frac{q_2^{(0)}}{(d-3)} \epsilon^2 + q_1^{(0)2} \frac{1}{2(d-3)} \epsilon^2 \right) \right] . \nonumber 
\end{eqnarray}
$S_P$ will have the same form, but with the sign of $q_1^{(0)}$ reversed, as the past surface is $t = - (z-\epsilon) + O(z^3)$. Thus, the correction to the action is 
\begin{eqnarray} 
S_F + S_P &=& 4 \frac{\ell^{d-1}}{\epsilon^{d-1}} \int_\Sigma d^{d-1} \sigma \sqrt{h} \left[ \ln (\sqrt{\alpha\beta} (d-1) \epsilon/\ell) ( 1 + q_0^{(2)} \epsilon^2) \right. \\ && \left. + \frac{1}{(d-1)} \left( 1 - q_0^{(2)} \frac{(d-1)}{(d-3)} \epsilon^2 - 2 \frac{q_2^{(0)}}{(d-3)} \epsilon^2 + q_1^{(0)2} \frac{1}{2(d-3)} \epsilon^2 \right) \right] .  \nonumber
\end{eqnarray}
Using the geometric expressions from \cite{Carmi:2016wjl},
\begin{equation} 
q_1^{(0)} = K, \quad q_2^{(0)} = \frac{1}{2} (K^2 + K_{ab} K^{ab} + R^a_a -R), \quad q_0^{(2)} = - \frac{1}{2(d-2)} (R^a_a - \frac{1}{2} R), 
\end{equation}
we can see that the logarithmic term will cancel with the contribution in \eqref{carmia}, as expected, including the subleading correction. The power law terms will combine with those in \eqref{carmia} to give us a result for the complexity
\begin{eqnarray} \label{oura}
\mathcal C_A(S) &=& \frac{\ell^{d-1}}{4\pi^2 G_N \epsilon^{d-1}} \int d^{d-1} \sigma \sqrt{h} \left[ \ln(d-1) \left( 1 - \frac{\epsilon^2}{2(d-2)} (R_a^a - \frac{1}{2} R) \right)  \right.   \\ && \left. -  \frac{\epsilon^2d}{2(d-1)(d-2) (d-3)} K^2 - \frac{\epsilon^2}{(d-2)(d-3)} K_{ab} K^{ab}  + \frac{\epsilon^2d}{2(d-1)(d-2)(d-3)} R   \right]. \nonumber
\end{eqnarray}
We see that this has a similar structure to the CV result \eqref{slcv}, but with different coefficients for the subleading terms.  

\subsection{Global AdS}

A simple example which illustrates the difference between CV and CA is to consider pure AdS in global coordinates,
\begin{equation} \label{glmet} 
ds^2 = \frac{\ell^2}{\cos^2 \theta} (-dt^2 + d\theta^2 + \sin^2 \theta d\Omega_{d-1}^2). 
\end{equation}
We consider a slice of the boundary at $t=0$, cutoff at $\theta = \theta_{cut}= \frac{\pi}{2} - \epsilon$. The maximal volume slice is again $t=0$ in the bulk, and the volume is simply 
\begin{equation} \label{gvol}
V(B) = \int d\theta d\Omega \sqrt{h} = \ell^d \Omega_{d-1} \int_0^{\theta_{cut}} \frac{d\theta}{\cos \theta} \tan^{d-1} \theta, 
\end{equation}
where $\Omega_{d-1}$ is the volume of a unit $S^{d-1}$.

\begin{figure}
\centering 
\includegraphics[width=3cm]{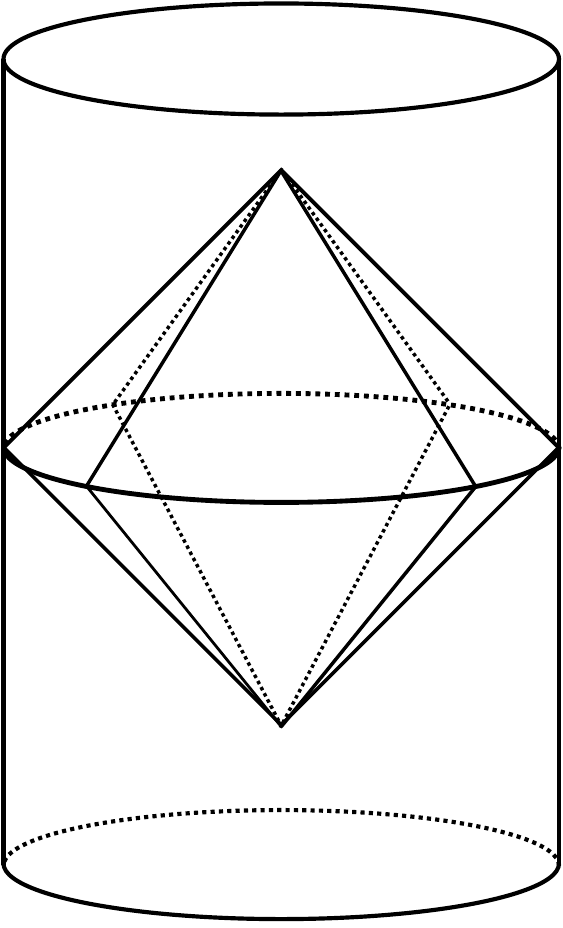}
\caption{The Wheeler-de Witt patch in global AdS.} \label{ads}
\end{figure}

We again calculate the action of the Wheeler-de Witt patch of the cutoff boundary at $\theta= \theta_{cut}$, as depicted in figure \ref{ads}. The future boundary is at $t= \theta_{cut} - \theta$, while the past boundary is at $t =  \theta - \theta_{cut}$. The volume term in \eqref{adsa} is 
\begin{eqnarray}
S_{\mathit{Vol}} &=& -\frac{4d}{\ell^2} \int_{0}^{\theta_{cut}} dt' \int_0^{t'} d\theta \int d\Omega_{d-1} \sqrt{-g}  \\ &=& - 4d\Omega_{d-1} \ell^{d-1} \int_{0}^{\theta_{cut}} dt' \int_0^{t'} d\theta \frac{\sin^{d-1} \theta}{\cos^{d+1} \theta} \nonumber \\ &=&  - 4\Omega_{d-1} \ell^{d-1}  \int_{0}^{\theta_{cut}} dt' \tan^d t'. \nonumber 
\end{eqnarray}
In the first step, we wrote the volume term as twice the integral over the future half of the Wheeler-de Witt patch.  We choose an affine parameter $\lambda$, so that the integrals over the future and past boundaries in \eqref{watact} do not contribute. An appropriate parameter is $\lambda = -\alpha^{-1} \ell \tan \theta$ on $F$ and $\lambda = \beta^{-1} \ell \tan \theta$ on $P$, where we have introduced the arbitrary parameters $\alpha$, $\beta$ purely so that we can see that they will cancel out once we add the term \eqref{acorr}. The future-pointing tangent is then
\begin{equation}
k = \frac{\alpha}{\ell} \cos^2 \theta (\partial_t - \partial_\theta) 
\end{equation}
on $F$ and 
\begin{equation}
\bar k = \frac{\beta}{\ell} \cos^2 \theta (\partial_t + \partial_\theta) 
\end{equation}
on $P$, so the boundary corner term is 
\begin{equation}
S_\Sigma = -2 \Omega_{d-1} \ell^{d-1} \tan^{d-1} \theta_{cut} \ln ( \alpha \beta \cos^2 \theta_{cut}).
\end{equation}
Adding the contribution \eqref{acorr},  the expansion on $F$ is 
\begin{equation}
\Theta = \frac{1}{\sqrt{\gamma}} \frac{\partial \sqrt{\gamma}}{\partial \lambda} = - \frac{1}{\sqrt{\gamma}}  \frac{\alpha}{\ell} \cos^2 \theta \frac{\partial \sqrt{\gamma}}{\partial \theta} = - \frac{\alpha}{\ell} \cos^2 \theta \cot^{d-1} \theta \partial_\theta (\tan^{d-1} \theta) = - \frac{ \alpha}{\ell} (d-1) \cot \theta, 
\end{equation}
and similarly on $P$ $\Theta = \beta /\ell (d-1) \cot \theta$. Thus the surface term is 
\begin{eqnarray}
S_F &=&- 2 \Omega_{d-1} \ell^{d-1} \int \Theta \ln |\Theta| \tan^{d-1} \theta d\lambda \\  &=& 2 (d-1) \Omega_{d-1} \ell^{d-1} \int_0^{\theta_{cut}} \frac{\tan^{d-2} \theta}{\cos^2 \theta}  \ln ( \alpha (d-1) \cot \theta) d \theta, \nonumber 
\\ &=& 2  \Omega_{d-1}  \ell^{d-1}  [ \tan^{d-1} \theta_{cut} \ln (\alpha (d-1) \cot \theta_{cut}) + \frac{1}{(d-1)}   \tan^{d-1} \theta_{cut}]  \nonumber
\end{eqnarray}
and similarly
\begin{eqnarray}
S_P& = &2 \Omega_{d-1} \ell^{d-1} \int \Theta \ln |\Theta| \tan^{d-1} \theta d\lambda\\ & = & \Omega_{d-1}  \ell^{d-1}  [ \tan^{d-1} \theta_{cut} \ln (\beta (d-1) \cot \theta_{cut}) + \frac{1}{(d-1)}   \tan^{d-1} \theta_{cut}]  
\nonumber
\end{eqnarray}
so in total
\begin{eqnarray} \label{adsfa} 
S &=&  S_V + \Delta S = S_{\mathit{Vol}} + S_\Sigma + S_F + S_P \\ &=& - 4\Omega_{d-1} \ell^{d-1}  \int_{0}^{\theta_{cut}} dt' \tan^{d} t' +4 \Omega_{d-1} \ell^{d-1}  \tan^{d-1} \theta_{cut} ( \ln(d-1) + \frac{1}{d-1}).\nonumber
\end{eqnarray}
We see that while the leading UV divergence is the same as for the volume \eqref{gvol}, the integrals are different, so the functional dependence on $\theta_{cut}$ is different for CV and CA. The two conjectures for the complexity are inequivalent. However, as noted in \cite{Carmi:2016wjl} (appendix C), the form of the subleading contributions in the CA calculation here depends on how we choose to cut off the Wheeler-de Witt patch, so it is not clear how much physical meaning it carries.

\section*{Acknowledgements}

AR is supported by an STFC studentship. SFR is supported in part by STFC under consolidated grant ST/L000407/1.

\bibliographystyle{JHEP}
\bibliography{complexity}

\end{document}